\documentclass[preprint2]{aastex}

\usepackage{natbib}

\usepackage{epstopdf}

\shorttitle{Formation of SiO grains at low temperature} 

\shortauthors{Krasnokutski et al.}

\begin{document}

\title{Formation of silicon oxide grains at low temperature}

\author{S. A. Krasnokutski, G. Rouill\'e, C. J\"ager, F. Huisken}
\affil{Laboratory Astrophysics Group of the Max Planck Institute for Astronomy
at the Friedrich Schiller University Jena, Institute of Solid State Physics, Helmholtzweg 3, D-07743 Jena, Germany}
\email{Sergiy.Krasnokutskiy@uni-jena.de}

\author{S. Zhukovska, Th. Henning}
\affil{Max Planck Institute for Astronomy, K\"onigstuhl 17, D-69117 Heidelberg, Germany}

\begin{abstract} 
The formation of grains in the interstellar medium, i.e., at low
temperature, has been proposed as a possibility to solve the
lifetime problem of cosmic dust. This process lacks a firm
experimental basis, which is the goal of this study. We have
investigated the condensation of SiO molecules at low temperature
using neon matrix and helium droplet isolation techniques. The
energies of SiO polymerization reactions have been determined
experimentally with a calorimetric method and theoretically with
calculations based on the density functional theory. The combined
experimental and theoretical values have revealed the formation of
cyclic (SiO)$_k$ ($k$ = 2--3) clusters inside helium droplets at $T$
= 0.37~K. Therefore, the oligomerization of SiO molecules is found
to be barrierless and is expected to be fast in the low-temperature
environment of the interstellar medium on the surface of dust
grains. The incorporation of numerous SiO molecules in helium
droplets leads to the formation of nanoscale amorphous SiO grains.
Similarly, the annealing and evaporation of SiO-doped Ne matrices
lead to the formation of solid amorphous SiO on the substrate. The
structure and composition of the grains were determined by infrared
absorption spectroscopy, transmission electron microscopy, and
energy-dispersive X-ray spectroscopy. Our results support the
hypothesis that interstellar silicates can be formed in the low 
temperature regions of the interstellar medium by accretion 
through barrierless reactions.
\end{abstract}


\section{INTRODUCTION}

Stars used to be seen as both the origin and the sinks of cosmic
dust grains. Indeed grains are formed in the shell of AGB stars and
supernovae. This so-called stardust is expelled from the stellar
environments, dispersed in the interstellar medium (ISM), and
eventually consumed by star formation processes. These processes
limit the lifetime of cosmic dust grains in the ISM to
$\sim$2$\times$10$^9$~yr \citep{Draine09}. If one takes additional
destructive mechanisms into account, e.g., shattering and sputtering
in interstellar shocks, their lifetime is further reduced to
$\sim$4$\times$10$^8$~yr \citep{Barlow77,Draine79,Jones94}. This
value, however, is not consistent with the timescale of dust
injection by stars of 3$\times$19$^9$~yr in the ISM. It was recently
estimated that only a few percent of the total mass of the dust in
the ISM is stardust \citep{Zhukovska08,Draine09}. This discrepancy
was first noted by \citet{Draine79} and later confirmed by several
other studies \citep[e.g.,][]{Dwek80,Weingartner99}. It suggests
that an intensive growth of dust particles actually takes place
directly in the ISM. At the present time it is assumed that most of
the material in interstellar dust grains was formed in the ISM
\citep{Weingartner99,Draine09}. However, \citet{Jones10}
have reconsidered the discrepancy between dust destruction and
formation processes in the ISM and found that based on inherent
uncertainties there could be a compatible injection and destruction
time scale of silicate dust.  They also noted that the spectra of grains 
where metals had been plated out on the surface
could be different from the spectra of silicates observed in the ISM.

Silicates are the main components of interstellar dust
\citep{Henning10}. The conditions for the silicate growth in the ISM
are presently not known. We assume that the grain growth occurs in
dark, dense molecular clouds at temperatures between 10 and 20~K
\citep{Herbst01}. Consequently, we intend to study the formation of
silicates at cryogenic temperatures. As a first step of our project,
we have carried out experiments on the condensation of SiO
molecules. The formation of (SiO)$_k$ ($k$ = 2--4) oligomers by
aggregation of SiO molecules was already found to proceed at low
temperatures, by means of laboratory experiments
\citep{Anderson68,Hastie69,Anderson69,Khanna81}. In addition,
quantum chemical calculations also predicted the barrierless
oligomerization reaction \citep{Lu03,Avramov05}. In contrast,
\citet{Pimentel06} determined an energy barrier of 8~kcal mol$^{-1}$
for the reaction between Si$_2$O$_2$ and SiO. Beside this initial
stage of cluster formation, first indications of the formation of
solid magnesium silicates were also observed when a nitrogen matrix
containing isolated SiO molecules and Mg atoms was evaporated. The
deposit produced in this way showed absorption bands at 10 and
20~$\mu$m, which were very similar to interstellar absorption
features \citep{Donn81}.

In this work, we extend these studies by addressing the  question of
energy barriers in the cluster formation processes and by evaluating
the binding energy of the clusters. Experiments have been carried
out in superfluid helium droplets at 0.37~K and in neon matrices at
6--11~K. They were completed with theoretical calculations. Finally,
we have analyzed the structure of grains formed at low temperature
in our experiments. We show that there is no barrier for the
aggregation of SiO molecules and that the grains formed in our
experiments are actually SiO grains with an amorphous structure and
a reactive surface.

\section{EXPERIMENTAL AND COMPUTATIONAL DETAILS}

\subsection{Matrix Isolation Spectroscopy} \label{exp:MIS}

The matrix isolation technique is a common method to study chemical
reactions at low temperatures and has been the subject of different
reviews \citep[e.g.,][]{Jacox02,Andrews04}. The matrix isolation
spectroscopy apparatus used in the present experiment was
extensively described in a previous publication \citep{Rouille12}.
Briefly, it consists of a commercial UV/VIS spectrometer (JASCO
V-670 EX) coupled to a vacuum chamber by means of optical fibers.
The chamber is equipped with a closed-cycle He cryocooler (Advanced
Research Systems Inc. DE-204SL). We used Ne (Linde, purity
99.995{\%}) as the matrix material. During the course of each
experiment, a CaF$_2$ substrate was placed into the vacuum chamber
and cooled to a temperature of 6~K by the action of the cryocooler.
Silicon monoxide molecules were prepared by laser vaporization
applied to silicon monoxide powder (Sigma-Aldrich) pressed into a
pellet. The fourth harmonic ($\lambda$ = 266~nm) of a pulsed ($f$ =
10~Hz) Nd:YAG laser was used for this purpose. Laser pulses of
1.5~mJ energy and 5~ns duration were focused to a spot of about
0.45~mm in diameter at the surface of the pellet. The evaporated
molecules were co-condensed onto the cold substrate with the rare
gas, which was fed into the vacuum chamber at a flow rate of 7~sccm
(standard cubic centimeter per minute).

\subsection{Helium Droplet Experiments} \label{exp:HeDroplet}

In spite of the many advantages provided by conventional solid rare
gas matrices, it has a serious limitation as the mobility of the
species trapped is considerably reduced. In contrast, liquid helium
droplets provide an ultralow temperature of $T$ = 0.37~K and are
proven to be superfluid \citep{Hartmann96b}. Therefore, all species
embedded inside the He droplets are allowed to move freely, thus
resembling the situation in the gas phase. As a result, helium
droplets provide an ideal possibility to study chemical reactions
and aggregations at low temperature. The doping of single helium
droplets with several species allows to study the condensation of
the embedded species. The formation of van der Waals clusters
\citep{Hartmann96a} and metal nanoparticles \citep{Loginov11a} was
demonstrated to occur inside the helium droplets. Alternatively, the
chemical reactions between embedded species can be studied at
ultralow temperature. The absence of an entrance channel barrier for
several chemical reactions was demonstrated
\citep{Krasnokutski10a,Krasnokutski10b,Krasnokutski11}. In
particular, the barrierless oxidation of silicon atoms and clusters
by oxygen molecules was found \citep{Krasnokutski10a}.

The He droplet apparatus is basically the same as that reported
earlier \citep{Krasnokutski05}. It consists of three differentially
pumped chambers. In the source chamber, large helium clusters are
produced by supersonic expansion of pure helium gas at high pressure
($p$ = 20~bar) through a 5~$\mu$m diameter pinhole nozzle, cooled by
liquid helium. The average number of He atoms ($N_\mathrm{He}$) per
droplet is evaluated according to the empirical relation
\begin{equation}\label{EqNrHe}
N_\mathrm{He} = 852393.161\,\mathrm{exp}\left(-0.3591\,T\right) ,
\end{equation}
where $T$ is the temperature of the nozzle in K. This equation has
been derived by fitting the experimental data of \citet{Toennies04}.
In the second chamber, the He droplets are doped with foreign
species. The doping occurs by collision of the helium droplets with
gas phase atoms or molecules. Upon the collisions, the gas phase
species are picked up by the helium nanodroplets and carried by them
to the third chamber. This chamber contains a quadrupole mass
spectrometer detector equipped with an electron bombardment ionizer.

Alternatively, an appropriate substrate can be placed into the
helium  droplet beam. In this case, after the collision of the
helium droplets with the substrate, the liquid helium is evaporated,
leaving the condensable content of the droplets on the substrate.
After removal of the substrate from the vacuum chamber, the
deposited material can be analyzed by high-resolution transmission
electron microscopy (HRTEM) and energy-dispersive X-ray spectroscopy
(EDX).

We performed four experiments using different  precursor
combinations (1. Si and H$_2$O molecules; 2. SiO molecules; 3. SiO
and H$_2$O molecules; 4. SiO, H$_2$O, and O$_2$ molecules) in helium
droplets. For the incorporation of Si atoms, commercially available
silicon wafer (CrysTec) was evaporated \citep[for details,
see][]{Krasnokutski10a}. In experiments with SiO molecules, a sample
of silicon monoxide (SiO) or synthetic quartz (SiO$_2$) powder was
heated in a zirconia crucible surrounded by a water-cooled jacket.
In both cases, SiO molecules are the main species released into the
gas phase. In case of the SiO$_2$ evaporation, additional oxygen is
produced. Silicon monoxide was used to study the initial cluster
formation by mass spectrometry while quartz was employed to condense
nanoscale grains. The vapor pressure of SiO decreases with
increasing evaporation time because of a structural modification of the
original SiO powder. The vapor pressure of SiO obtained by heating
quartz powder is more stable, but requires a higher evaporation
temperature. Thus silicon monoxide powder is more suitable as a
precursor for the fast mass spectrometry measurements while quartz
powder is chosen for the more time-consuming study of grain
condensation. Finally, it is not necessary to use a dedicated source
to incorporate H$_2$O molecules in He droplets as H$_2$O is present
in every standard vacuum apparatus as residual gas. In studies
involving water molecules, the residual gas pressure was adjusted to
a stable value of approximately 4$\times$10$^{-6}$~mbar. When the
incorporation of water was not desired, the concentration of water
molecules was reduced by filling a cold trap with liquid nitrogen,
which resulted in an improvement of the pressure to about
3$\times$10$^{-7}$~mbar. At this condition, only a minor
amount of helium droplets (less than 3 \%) picked up the water
molecules. Due to the geometry of the apparatus, the pickup of water
molecules occurs after that of the Si or SiO species. The
concentration of water and other impurities was monitored by
quadrupole mass spectrometer analyses. The incorporation of other
impurities into the He droplets could not be detected.

\subsection{Theoretical calculations} \label{exp:TheoCalc}

Molecular geometries of (SiO)$_k$ clusters were determined
using the B3LYP/6-311+G** and CCSD(T)/cc-pVTZ levels of theory
implemented in the GAUSSIAN09 software package \citep{Frisch09}.
The reaction energies were obtained as the difference between the
electronic and zero point energies of the reactant and product
molecules. CCSD(T)/cc-pVTZ quantum chemical calculations 
provide high levels of accuracy of about 20~kJ mol$^{-1}$ for the reaction
energy \citep{Delley06}.

\section{RESULTS}

\subsection{Matrix Isolation Study} \label{res:MIS}

\begin{figure}[b!]
\epsscale{0.80} \plotone{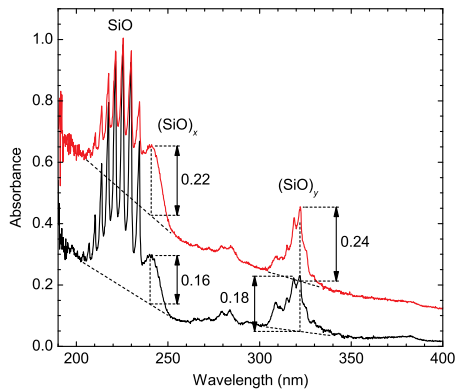} \caption{Electronic spectra of single SiO molecules and their oligomers isolated in a Ne matrix at $T$ = 6~K. The lower (black) spectrum was obtained directly after co-deposition of the laser-vaporized SiO molecules and the matrix gas. The upper (red) spectrum shows the result of an annealing of the matrix for 15~min at $T$ = 10~K and 10~min at $T$ = 11~K. (A color version of this figure is available in the online journal.)\label{fig1}}
\end{figure}

The laser-evaporated SiO molecules were co-condensed with the  Ne
gas on the CaF$_2$ substrate at 6~K. The UV spectrum of this matrix
is displayed in Figure~\ref{fig1}. The vibrational progression of
the $A^1\Pi \leftarrow X^1\Sigma^+$ transition of SiO was previously
observed in other inert matrices \citep{Shirk68,Hormes83}  and can
be easily identified in our spectrum where it spreads from 235~nm
towards shorter wavelengths. In addition, one observes a broad band
at 240~nm and two structured features at 280 and 320~nm. Then, we
increased the temperature of the substrate several times to $T$ =
10--11~K. At these temperatures, the Ne matrix slowly starts to
sublimate. Additionally, all isolated species gain some partial
mobility inside the matrix, which allows these species to aggregate
and form clusters. As can be seen in Figure~\ref{fig1}, after an
annealing at 10 and 11~K, the intensity of the SiO bands was
noticeably reduced. This is due to the aggregation of isolated SiO
molecules and the formation of some larger clusters. At the same
time, the absorption features at 240 and 320~nm increased in
intensity. Therefore, these bands are likely due to (SiO)$_k$
clusters. This assignment is also supported by the fact that, in
previous studies, (SiO)$_2$ and (SiO)$_3$ molecules were present in
the matrices together with SiO and that the intensity of the
absorption bands assigned to these molecules rose upon annealing of
the matrix \citep[and references therein]{Friesen99}. Moreover, the
absorption features of these molecules were predicted to be in the
range between 200 and 400~nm \citep{Reber08,Xu10}. Our results are
in line with the previous studies of SiO aggregation at low
temperatures, which demonstrated the formation of small (SiO)$_k$
($k$ = 2--4) clusters using different experimental techniques
\citep{Hastie69,Anderson69,Khanna81,Schnoeckel89,Friesen99,Pimentel06}.

\begin{figure}[t]
\epsscale{0.50} \plotone{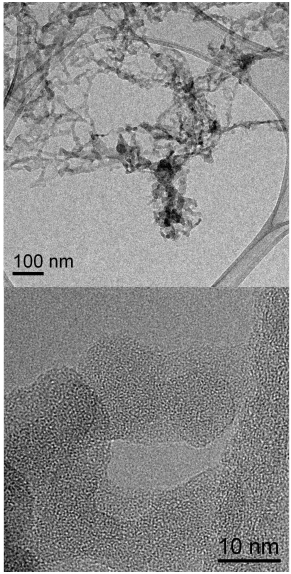} \caption{TEM images showing the
condensate collected after the annealing and evaporation of a Ne
matrix doped with SiO molecules and oligomers.\label{fig2}}
\end{figure}

After having taken the spectra of Figure~\ref{fig1}, the
temperature of the matrix was slowly raised, allowing the matrix to
evaporate completely. After the evaporation of the matrix, a solid
condensate remained on the substrate. In situ UV spectra
measurements, taken at room temperature while the substrate was
still under vacuum, have demonstrated that the extinction caused by
the condensate was characterized by a smooth curve rising towards
shorter wavelengths. Therefore, the observed extinction can be
attributed to scattering caused by the grains. Their analysis by EDX
revealed a Si/O ratio of about 1. Transmission electron microscopy
(TEM) images of the particles are shown in Figure~\ref{fig2}. All
observed particles have an amorphous structure with the size of the
smallest grains being around several tens of nanometers. The smaller
particles form larger aggregates of micrometer size.

\begin{figure}[t]
\epsscale{0.80} \plotone{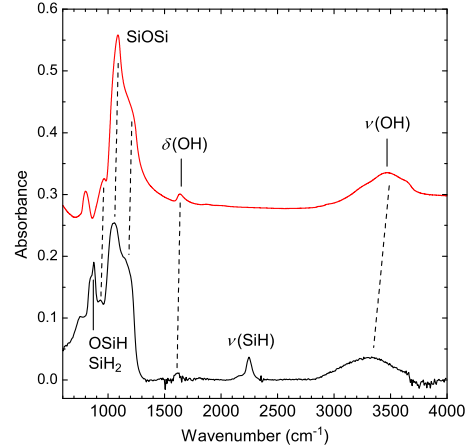} \caption{Infrared spectrum of a
single 20~$\mu$m particle collected after the annealing and
evaporation of a Ne matrix doped with SiO molecules and oligomers
(lower curve, black) in comparison to a SiO$_2$ material produced by the
sol gel technique (upper curve, red). (A color version of this figure is available in the online journal.) \label{fig3}}
\end{figure}

A single grain of about 20~$\mu$m in diameter was selected from the
condensate to measure its infrared (IR) absorption spectrum in a
range between 600 and 4000~cm$^{-1}$ using an IR microscope. The
spectrum is shown in Figure~\ref{fig3}. Absorption peaks are
observed at 751, 843, 873, 932, 1052, 1164, 1614, 2159, 2248, and
3327~cm$^{-1}$. The strong bands at 1052 and 1164~cm$^{-1}$ are
assigned to the transverse and longitudinal modes of the Si--O--Si
asymmetric stretching vibrations commonly observed in SiO$_2$.
For the purpose of comparison, a spectrum of amorphous
SiO$_2$ produced by the sol gel technique was added to
Figure~\ref{fig3}. The strong similarity of the broad 10~$\mu$m band
with the band of a SiO$_2$ material point to a certain
stoichiometric separation and disproportionation of the SiO into
silicon and silica. Recent experimental studies on the structure and
chemical composition of solid SiO clearly confirm the inhomogeneity
of solid SiO in a nanometer- or subnanometer scale. The formation of
phases consisting of Si and SiO$_2$ and the formation of the typical
SiO$_4$ groups connected by oxygen bridges in addition to Si-Si
groups have been proven by different authors
\citep{Ferguson12,Hohl03}. In addition, quantum chemical
computations comparably demonstrate the tendency of SiO clusters to
form SiO$_4$ in addition to Si-dominated groups
\citep{Reber08,Goumans13}.

Beside these bands, the spectrum displays peaks that are not  found
in the commercial silicon monoxide compound, which was measured for
comparison. The peaks at 2248 and 3327~cm$^{-1}$ are attributed to
the stretching vibrations of SiH and OH groups, respectively. These
groups were possibly formed in the condensation or by reaction of
the freshly-condensed SiO grains with residual H$_2$O molecules
during the warming to room temperature in the vacuum chamber.
However, the OH groups can also be due to water molecules adsorbed
on the surface of the grains and of the KBr substrate. The formation
of silanol groups (Si--OH) either on the surface or within the
silicate structure during the condensation and the warm up of the
substrate is very likely. The presence of a small Si-OH stretching
band at 932~cm$^{-1}$ confirms this assumption.

The OH stretching bands of single, paired, and hydrogen-bonded
silanol groups are expected between 3750 and 3540~cm$^{-1}$
\citep{Tyler69,Armistead69}. Due to the presence of water vapor
lines near 3800~cm$^{-1}$, it is difficult to ascertain the presence
of a band corresponding to OH stretching modes of such silanol
groups. The broad band at 3327~cm$^{-1}$ that can tentatively be
assigned to stretching vibrations of interacting water molecules and
silanol groups (Si--OH) \citep{Zhdanov87} supports the existence of
these groups. However, a firm discrimination between water molecules
either adsorbed on the surface of the grains or on the KBr is
difficult. The weak band at 1614~cm$^{-1}$ is attributed to the
corresponding bending mode $\delta$(OH) of water molecules whereas
bands at 843 and 873~cm$^{-1}$ can be assigned either to vibrations
of O-SiH groups \citep{Shinoda06} or to vibrations of Si--Si bonds
of the sesquioxide group Si$_2$O$_3$ \citep{Day78,Hagenmayer98}.
A similar band was already observed in amorphous silicates
of olivine and pyroxene stoichiometry produced by condensation in a
H$_2$O containing atmosphere at high temperatures \citep{Jaeger09}.
Our experiments have shown that Si-H bonds can easily be destroyed 
by oxygen or oxygen-bearing molecules. In the dense ISM, 
where a lot of oxygen-bearing molecules such as H$_2$O, CO, 
CO$_2$, and many other species plate out on the surface of the cold grains, reactions between Si-H bonds and these molecules may occur resulting in 
the formation of Si-O bonds. Of course, we do not exactly know the energy barriers for these reactions, but cosmic rays may help to speed up the processes. In less dense regions of molecular clouds and in diffuse cold clouds, 
the substitution of H can be enhanced by UV irradiation.
All things considered, the IR spectral measurements of the
condensate clearly show the formation of a solid SiO$_x$ compound
that reveals the spectral signatures of SiO$_2$.

\subsection{Helium Droplet Isolation Studies} \label{res:HeDroplet}

To study the condensation of silicon oxide grains inside
helium droplets, we apply two different approaches. We study the
co-condensation of water molecules together with Si atoms and
together with SiO molecules. Both routes should be relevant to the
processes occurring in the ISM. Depending on the conditions for the
destruction of the stellar silicate grains, the formation of SiO
molecules or Si atoms is expected. Additionally, Si and SiO species
may arrive to the ISM directly from the stellar regions, and SiO
molecules can be formed in the reaction Si + (O$_2$, OH)
\citep{Turner98}. At the same time, the water molecule is the most
abundant oxygen-bearing molecule in the Universe and could serve as
an oxidizer for Si and SiO species, which are accreted on the
surface of dust particles.

\subsubsection{Initial Cluster Formation} \label{res:HeIniStage}

The initial stage of cluster formation was studied by the means of
quadrupole mass spectrometry. The mass spectra were obtained after
electron impact ionization of the doped helium droplets. The
ionization of a helium droplet results in the formation of He$^+$.
This positive charge migrates through the droplet via a
positive-hole resonant-hopping mechanism. In the case of relatively
small helium droplets with less than 30\,000 helium atoms, the
migration process terminates when the charge is transferred from
He$^+$ to the dopant or when the charge becomes localized at another
helium atom to create He$_{n}^{+}$ \citep{Scheidemann93}. The excess
energy ejects the ionized dopant or the He$_{n}^{+}$ cluster from
the droplet. As a result, the dopant and helium cluster cations are
observed as fragments of the helium droplets in the mass spectrum on
their respective masses. Unfortunately, with the increase of the
helium droplet size, the probability of charge transfer to the
dopant is reduced and only He$_{n}$ peaks can be detected by mass
spectrometry. Therefore, mass spectrometry combined with electron
impact ionization cannot be applied to study the formation of large
clusters as their assembly requires much larger helium droplets
containing a few millions of helium atoms.

\begin{figure}[th]
\epsscale{0.80} \plotone{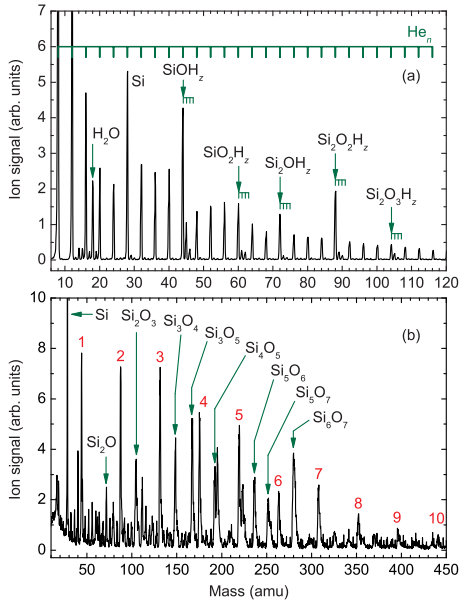} \caption{Mass spectra of He droplets doped with Si atoms and H$_2$O molecules (a) and with SiO and H$_2$O molecules (b). In panel (a), Si$_x$O$_y$H$_z$ compounds are indicated with (green) arrows. The small scales beneath the arrowheads mark the different values of $z$ ($z$ = 0--3). In panel (b), the Si$_x$O$_y$H$_z$ compounds with $z$ $>$ 0 are not resolved, therefore they are not indicated. The mass peaks corresponding to (SiO)$_k$ clusters are labelled (in red) with values of the natural number $k$. (A color version of this figure is available in the online journal.)\label{fig4}}
\end{figure}

Figure~\ref{fig4}(a) shows the mass spectrum of helium droplets
doped with silicon atoms and water molecules. Compounds with the
formula Si$_x$O$_y$H$_z$ ($x$ = 1--2, $y$ = 1--3, and $z$ = 0--3)
can be identified. Although it is not shown with this figure, the
peak intensities of products with larger $y$ and $z$ grow as the
water concentration increases. As it was previously found that Si
atoms and small clusters do not form chemical bonds with H$_2$O
molecules inside He droplets \citep{Krasnokutski10a}, these
compounds were formed by ion-molecule reactions after charge
transfer from He$^+$ to the Si$_x$(H$_2$O)$_y$ complexes.

Figure~\ref{fig4}(b) depicts the mass spectrum measured when the He
droplets are doped with SiO and H$_2$O molecules. This mass spectrum
was obtained with the lower-resolution mode of the mass
spectrometer, which allows to access a considerably higher mass
range. Mass peaks corresponding to various Si$_x$O$_y$ clusters can
be observed, including (SiO)$_k$ clusters indicated with values of
$k$. The observation of (SiO)$_k$ clusters formed in He droplets is
consistent with the results of our matrix isolation study (see
Section \ref{res:MIS}) and with literature data (see Introduction).
We suppose that the presence of species with higher oxygen content
($y$ $>$ $x$) reveals ion-molecule reactions between SiO and H$_2$O
species rather than the loss of Si atoms upon ionization. The low
resolution of the spectrum does not allow us to clearly separate the
Si$_x$O$_y$H$_z$ from the Si$_x$O$_y$ species, however, the presence
of such cations is firmly determined. Specific mass peaks exhibit
much higher intensities than other peaks in the same mass range. The
large intensity of these mass peaks reflects the high stability of
the corresponding cations and reveals a considerable fragmentation
caused by the ionization or the reactions triggered by the
ionization process.

To study the reactions of SiO molecules inside the droplets
before the ionization point, we applied the calorimetric approach.
In the case of a chemical reaction, the energy released during the
reaction is transferred to the He droplet, causing the evaporation
of He atoms. The change of the He droplet size upon reaction can be
detected in the experiment. This calorimetric technique was already
applied to study chemical reactions inside helium droplets
\citep{Krasnokutski10a,Krasnokutski11}. Another calorimetric method
recently developed by \citet{Lewis12} is well suited to the study of
cluster formation. The idea behind this technique is the
determination of the smallest helium droplet size that allows the
assembly of the desired cluster inside the droplet. If the helium
droplet is smaller than this minimum size it would be completely
destroyed upon the reaction and it would not arrive at the detector.
As a result, during a continuous increase of the helium droplet size
by varying the droplet source temperature, a distinct rise of the
number of (SiO)$_k$ cations occurs when the helium droplet becomes
larger than the minimum size required for the dissipation of the
binding energy of this (SiO)$_k$ cluster. We have applied the basic
principle of this technique, with a development of our own, to
monitor the formation of SiO clusters. The detailed description of
the procedure is given in Appendix~\ref{app1}.

\begin{table*}[t]
\begin{center}
\caption{Reaction Energies\label{table1}}
\begin{tabular}{lccccc}
\tableline
\tableline
Reaction & \multicolumn{5}{c}{Energy} \\
 & Observed & \multicolumn{4}{c}{Theoretical} \\
 & & Model 1\tablenotemark{a} & Model 2\tablenotemark{b} & Model 3\tablenotemark{c} & Model 4\tablenotemark{d} \\
\tableline
SiO + SiO $\rightarrow$ Si$_2$O$_2$                          & 178 & 170 & 211 & 201 & 163 \\
Si$_2$O$_2$ + SiO $\rightarrow$ Si$_3$O$_3$                  & 291 & 248 & 240 & 241 & 253 \\
(SiO + SiO) + H$_2$O $\rightarrow$ Si$_2$O$_2$OH$_2$         & 206 & 172 &     &     &     \\
(Si$_2$O$_2$ + SiO) + H$_2$O $\rightarrow$ Si$_3$O$_3$OH$_2$ & 249 & 244 &     &     &     \\
\tableline
\end{tabular}
\tablenotetext{a}{B3LYP/6-311+G**, this work.}
\tablenotetext{b}{CCSD(T)/cc-pVTZ, this work.}
\tablenotetext{c}{B3LYP/6-31G** \citep{Friesen99}.}
\tablenotetext{d}{QCISD/6-311++G** \citep{Pimentel06}. The original
values were expressed in units of kcal mol$^{-1}$.}
\tablecomments{Reaction energies are expressed in units of kJ
mol$^{-1}$.}
\end{center}
\end{table*}

The reaction energies we have determined experimentally are
summarized in Table~\ref{table1} together with the values obtained
from quantum chemical calculations. All computational values
presented in the table were derived assuming cyclic reaction
products. They are consistent with our experimentally derived values
for binding energies that confirm the formation of cyclic products
by barrierless reactions.
This is in agreement with the results of previous IR spectroscopy
studies of cryogenic matrices
\citep{Anderson68,Hastie69,Anderson69}. However, \citet{Friesen99}
did not observe barrierless reactions for the SiO oligomerization in
methane cryogenic matrices. Moreover, \citet{Pimentel06} determined
an energy barrier of 8~kcal mol$^{-1}$ for the reaction between
Si$_2$O$_2$ and SiO.

At the current state, it is difficult to accurately
calculate the error for the experimental determination of the
reaction enthalpy within the helium droplets since a few
simplifications were used in the analysis. In Appendix~\ref{app1}, a more
detailed discussion of the error is performed and it is estimated
to be smaller than 30\%. Therefore,the difference between the
present computational and experimental values is within the
uncertainty interval of the measurements and thus suggests that
these reactions indeed took place inside the helium droplets at $T$
= 0.37~K and led to the formation of cyclic (SiO)$_k$ clusters. This
demonstrates that the formation of cyclic (SiO)$_k$ clusters is
barrierless and should be fast in the entire low-temperature range
and its reaction rate is defined by the collision probabilities of
the SiO molecules. The absence of an energy barrier for the
insertion of the SiO molecule into the Si$_2$O$_2$ ring with 
four equally strong Si-O bonds demonstrates
that the existing Si--O bonds can easily rearrange into
energetically favored structures. Although Si$_x$O$_y$ compounds
linked to the presence of H$_2$O were observed in the experiments,
we suppose that they were formed by the ionization process.

\subsubsection{Grain Formation} \label{res:HeGrain}

\begin{figure*}[t]
\epsscale{1.5} \plotone{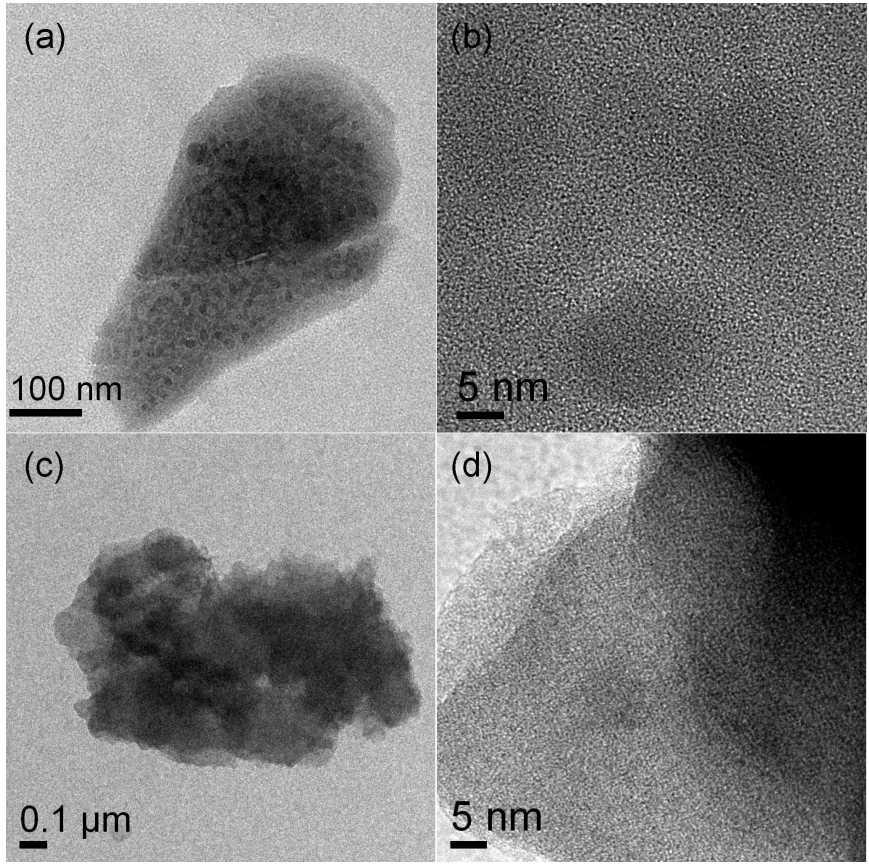} \caption{TEM images showing the particles formed on the substrate placed in the He droplet beam when the droplets were doped with Si atoms and H$_2$O molecules [(a) and (b)], and with SiO and H$_2$O molecules [(c) and (d)].\label{fig5}}
\end{figure*}

To produce solid material inside the droplets, the nozzle
temperature was reduced to $T$ = 8~K resulting in He droplets
containing about 10$^7$ He atoms. The vapor pressure of the dopants
was kept the same as in the previous experiments. Nevertheless, the
40-times increase of the geometrical cross section of the He
droplets also resulted in an increase of the number of dopants
picked up. To collect the condensate formed in the droplets, a lacey
carbon TEM grid was inserted into the beam of the doped helium
droplets. The deposit on this grid was analyzed by HRTEM and EDX.
Figure~\ref{fig5} shows HRTEM images of the deposit obtained with
helium droplets doped with Si atoms and H$_2$O molecules (1.
experiment) [Figures~\ref{fig5}(a) and \ref{fig5}(b)] and with SiO,
O$_2$, and H$_2$O molecules (4. experiment) [Figures~\ref{fig5}(c)
and \ref{fig5}(d)]. In both cases, we observe large silicon oxide
grains, which cannot be produced inside single helium droplets. This
demonstrates that additional aggregation of the nanometer-sized
condensates takes place on the substrate.

The TEM image in Figure~\ref{fig5}(a) shows numerous dark spots
inside a grain produced by doping helium droplets with Si and
H$_2$O. The EDX analysis of these spots has demonstrated their
chemical identity with the main body of the particle, pointing to a
thickness contrast, exclusively. Therefore, these spots likely
represent particles grown in individual He droplets. Additional
support for this interpretation is provided by the fact that the
size of these smaller particles is about 10~nm [see
Figure~\ref{fig5}(b)], which is also the size of the silver
nanoparticles grown in helium droplets under similar experimental
conditions \citep{Loginov11b}.

The analysis of the TEM image in Figure~\ref{fig5}(c) yields a
similar conclusion concerning condensation experiments with SiO and
H$_2$O/oxygen. The slowly decreasing SiO vapor pressure, however,
leads to the formation of smaller particles in the helium droplets,
with an average size of 3~nm, as shown in Figure~\ref{fig5}(d). The
EDX analysis revealed the formation of SiO$_2$ grains that supports
the concept of a condensation within the He clusters since a later
oxidation of SiO grains with oxygen molecules at room temperature is
found to be very slow \citep{Fogarassy87}.

\section{ASTROPHYSICAL DISCUSSION}

In our experiments, we have observed the formation of silicon oxide
grains at low temperature. In the matrix isolation experiment, the
SiO grains were formed by reaction between SiO molecules during the
annealing (10--11~K) and evaporation of the Ne matrix (12~K), that
is under conditions comparable to those expected in molecular
clouds. The study of different dopant molecules (Si and H$_2$O; 
SiO; SiO and H$_2$O; SiO, H$_2$O, and O$_2$) in superfluid He droplets has shown that the reaction between SiO molecules can take place already at 0.37~K and
is therefore barrierless. Consequently, the condensation of solids
should be as well possible at higher temperatures (10--300~K). Furthermore,
the results have shown that Si atoms can be easily oxidized.

Certainly, solid SiO is not considered as an abundant cosmic dust
component. Pure SiO$_x$ grains are not observed in the ISM while the
presence of Mg-Fe silicates has been proven \citep{Henning10}.
However, gaseous SiO is one of the main precursors for silicates in
evolved stars \citep{Gail99}. 

In the ISM, SiO can be produced by different ways, such as photodesorption 
of SiO from silicate grains by UV irradiation \citep{Turner98}, 
and sputtering of Si followed by the oxydation to SiO molecules in shock waves
\citep{Gusdorf08a,Gusdorf08b}. SiO emissions in molecular clouds because of
shock waves have also been observed \citep{Dishoek93,Martin97,Handa06,Ngyuen13}.
Recently, the formation of pure SiO grains in the shells of S-stars has been
proposed \citep{Wetzel13}. The study of the formation of SiO
condensates at low temperatures is the first step into the direction
of the more complex magnesium-iron silicates of olivine and pyroxene
stoichiometry.

The dust condensation at low temperature and low density in the ISM is 
a process that is discussed to be necessary to keep the balance between 
dust destruction and formation \citep{Chiaki13}. However, there is no exact description where the cold condensation process may finally take place. 
Dust can be easily destroyed in SN shocks. Turbulence 
can quickly distribute the delivered refractory elements such as Mg, Si, 
and other species into the surrounding ISM. \citet{Oey03} estimated that 
freshly synthesized metals from SN ejecta are mixed by turbulence with the 
surrounding ISM on a time scale of 100 Myr. One can assume a similar time 
scale for the mixing of elements liberated from destroyed grains.

The process of cold condensation may already be active in diffuse clouds with low temperature and in low-density regions of molecular clouds. In molecular cloud environments with maximum density, where many 
non-refractory and refractory species and elements may accrete simultaneously on the grains, rather complex ices are formed and the formation of silicates and other solids could be restricted.
However, dust grain formation is essentially a process which can be described
by two steps: nucleation and growth. If the elements which are
heavily depleted in the ISM stick efficiently to the surviving
grains, then, accretion onto grains occurs quickly enough to account
for their depletion. A previous nucleation step is not necessary to
form solids. If refractory atoms and molecules accrete slowly onto
the surfaces of surviving grains, such atoms have the opportunity to
react with additional adsorbed atomic and molecular species before
being buried beneath a growing layer. The most abundant atomic
species is H and the most abundant molecules are H$_2$, CO, and
H$_2$O. Therefore, reactions between refractory elements and these
molecules might also form complex ices. There are two competitive 
processes affecting the growth of the layer which is the desorption 
of molecules due to irradiation and the sticking. While the sticking of 
an atom or molecule to a grain is a process of low selectivity at low 
temperatures, the desorption process is strongly influenced by the bonding 
energy between the atom or molecule and the surface. According to \citet{Draine09}, 
for binding energies of 0.05~eV, the lifetime against thermal desorption is 
about 1~s, whereas for a binding energy of 0.1~eV, the lifetime is increased 
to a value of 5$\times$10$^5$~yr. Molecules or atoms that may form strong bonds,
which are typical for refractory solids, grow very fast and remain on the
surface for a long time. Moreover, diffusion/desorption processes on grain 
surfaces may occur, and eventually, lead to the formation of stable solids 
such as silicates or carbonaceous material in addition to complex ices.
Such possible selection processes have to be addressed in upcoming experimental studies.

In addition, our experiments have shown that SiO can form clusters
that can finally condense as solid silicon oxide. Since these reactions
occur without an energy barrier, no further irradiation with UV or
cosmic-ray energy protons is necessary to form solids. 
Barrierless structural rearrangements in SiO clusters are possible. 
This has been demonstrated by quantum chemical computations that 
show the tendency of  SiO clusters to disproportionate 
into SiO$_2$ and Si phases \citep[Bromley~priv.~communication]{Reber08,Goumans13}. 
UV irradiation may help to rearrange the bonds in the SiO clusters. 
Furthermore, UV irradiation can help to desorb non-refractive species that 
form complex ices and prevent the formation of solids. 
Moreover, all gaseous species should be depleted on grains within a 
timescale of 1$\times$10$^5$~yr in dense clouds. Since these clouds have 
longer lifetimes of about 10~Myr and some species are still observed in 
the gas phase, there must be a mechanism to frequently remove the mantles 
of accreted species from the grain surfaces. Such mechanisms should result 
in the preferred formation of compounds with higher binding energy such as silicates
and should remove the weakly bound species in ice mantles.

In general, dense molecular clouds are not completely protected from
the interstellar UV field since diffuse UV photons from outside and
from newly born stars inside the clouds can deeply penetrate even
the interior of such a cloud. In addition, cosmic rays can penetrate
the molecular clouds and similarly trigger reactions between
accreted molecules and clusters. Previous studies of reactions in He
clusters showed that Si, Mg, and Al atoms also reacted with O$_2$
molecules at very low temperatures
\citep{Krasnokutski10a,Krasnokutski10b,Krasnokutski11}. Of course,
the fact that there is no energy barrier at the entrance channel of
a reaction between two atoms or molecules does not mean that its
product can be finally formed in the gas phase in the ISM. Actually,
the excess energy generated by the reaction can cause the
dissociation of the product if it cannot be released by another
means like the emission of photons. When a species reacts with the
surface of a preexisting grain, however, the excess energy can be
transferred to this grain, which plays the role of a heat sink.

In the ISM, UV irradiation with 10~eV photons and low-energy ion
irradiation due to acceleration of ions and atoms in shock waves can
strongly force the formation of solids. However, polymerization and
photodesorption of molecules are competitive processes.  Chemical
reactions between adsorbed molecules and the grain surface can
prevent the desorption of the adsorbate. In the grains condensed at
low temperatures, the defect formation is a very probable process
since the diffusion of atoms within the structure of the solids is
strongly slowed down preventing the annealing of possible defects.
In addition, IR spectroscopy of the condensed grains clearly show
the appearance of IR bands at 873 and 2248~cm$^{-1}$ which point to
the presence of Si--H bonds or special structures related to the
sesquioxide group Si$_2$O$_3$. These bands appear in addition to the
two typical bands caused by Si--O bonds in silicon oxides or in
silicates. Si--Si or Si--H bonds that can be easily destroyed upon
UV or ion irradiation can finally act as such defects and may change
the sticking behavior of the molecules on the grain surface.

Experimental studies on the carbon dust formation in the ISM
performed by \citet{Jenniskens93} have shown the formation of carbon
dust upon UV and ion irradiation. The authors found that dust
particles accrete a layer of about 20~nm during their lifetime in a
molecular cloud. UV photons clearly dominate the polymerization and
carbonization process even though the process is much more efficient
for ion-induced polymerization.

The SiO grains formed by low temperature condensation have an
irregular, fluffy morphology. In that respect they do not differ
from grains condensed at relatively higher temperatures
\citep{Colangeli03}. Their structure is amorphous and therefore very
well comparable to the structure of the observed interstellar
silicates. The degree of crystallinity for interstellar
silicates was determined not to be higher than 2.2\%
 \citep{Kemper04,Kemper05,Min07}.

Future experiments on dust formation in the ISM including SiO and
refractory elements such as Mg and Fe in cryogenic matrices and on
bare surfaces are in progress. In situ IR spectroscopy of the condensed
layers during the growth process is necessary to follow the
formation of bonds and defects and the effect of annealing in the
structure of the condensate.

\section{CONCLUSION}

The condensation of silicon monoxide at low temperature was  studied
in solid neon matrices and superfluid helium droplets. The initial
stage of condensation was investigated in the helium droplets by
mass spectrometry and by quantum chemical calculations. The
formation of cyclic (SiO)$_k$ ($k$ = 2--3) clusters inside helium
droplets at $T$ = 0.37~K has been found to be barrierless. The
binding energies of (SiO)$_k$ ($k$ = 2--3) clusters were obtained by
measuring the minimum size of helium droplets that is necessary for
assembling the desired cluster. Additionally, the formation of
nanometer to micrometer-sized silica grains was observed after
evaporation of the liquid helium droplets or the neon matrix doped
by single SiO molecules. These results clearly demonstrate
the formation of solid SiO at low temperature, which is an important 
requirement for the formation of more complex silicates in the entire low
temperature range of the ISM. The analysis of the particles reveals
an amorphous structure with a high surface area. Thus it appears
necessary to consider cosmic dust analogs formed at low temperatures
when interpreting astrophysical observations. The data presented in
this article demonstrate the power of the matrix and helium droplet
isolation techniques to study low temperature dust formation. The
extension of this study to silicates with different Mg and Fe
contents and with in situ spectral characterization of the dust
grains formed in the experiments, is planned.

\acknowledgements This work was supported by the Deutsche
Forschungsgemeinschaft, DFG project number He 1935/26-1, which is a 
part of the DFG priority program 1573 'Physical Processes in the ISM'.
We are grateful to Frank H\"anschke, 
IPHT Jena, for performing the FTIR microscope measurement.
We are also grateful to J. Nuth for his helpful discussions.

\appendix
\section{EVALUATION OF REACTION ENERGIES} \label{app1}

In our He droplet experiment, the vapor pressure of SiO was  kept
constant and the number of cations at specific masses was counted as
a function of the helium droplet size. We monitored the ion
count rate at the masses of (SiO)$_k$ clusters and their complexes
with water molecules (SiO)$_k$OH$_2$. In the latter case, we set
the vapor pressure of water so as to achieve a probability of doping
the helium droplets with H$_2$O molecules close to 100{\%}. As a
result, the signal at the mass of any (SiO)$_k$OH$_2$ compound is
defined by the number of helium droplets containing the
corresponding (SiO)$_k$ cluster. Monitoring  the (SiO)$_k$OH$_2$
signal has the advantage that the mass of such a compound does not
coincide with the mass of a He$_n$ cluster, unlike that of a
(SiO)$_k$ cluster. Additionally, these complexes can be formed only
inside the He droplets.

\begin{figure}[b!]
\epsscale{0.50} \plotone{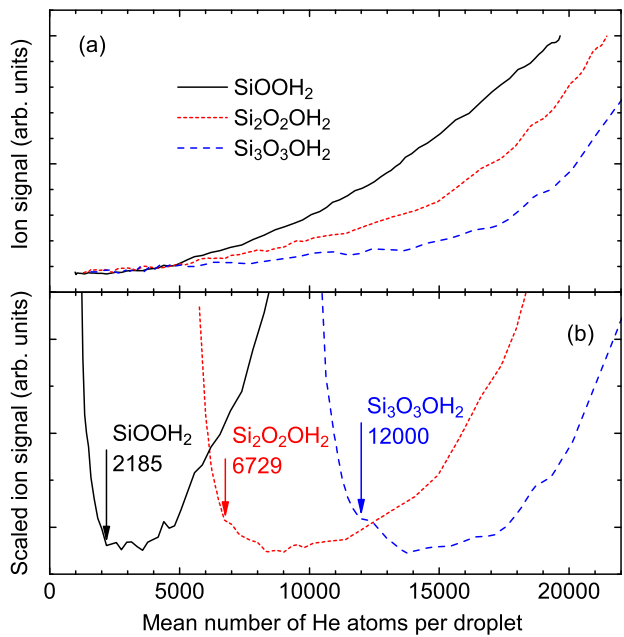} \caption{Mass spectrometry signal of various cations as a function of the average number of He atoms per droplet. Panel (a) shows the raw measurements. The data depicted in panel (b) have been derived by scaling the data of panel (a). (A color version of this figure is available in the online journal.)\label{fig6}}
\end{figure}

Ion count rates as a function of the helium droplet size are
shown in Figure~\ref{fig6}(a). They increase with the size of the
helium droplets, reflecting the fact that larger droplets pick up
more molecules and they are more efficiently ionized. The onset of
the ion signal rise for increasing (SiO)$_k$ cluster sizes is
shifted towards bigger helium droplets revealing the larger amount
of energy released in the oligomerization processes. In contrast with
previous studies by \citet{Lewis12}, our ion signals do not show
sharp thresholds. Their intensities rise rather smoothly towards
larger helium droplet sizes. This observation can be understood
taking into account that the sizes of the He droplets follow a
log-normal distribution. To find the minimum helium droplet
size required for the (SiO)$_k$ cluster formation, we compared the
experimentally derived numbers of (SiO)$_k$ ions with the calculated
probability of (SiO)$_k$ ion formation $P$  that can be calculated
with
\begin{equation}\label{eqP}
P=P_{k}P_\mathrm{I}P_\mathrm{CT} ,
\end{equation}
where $P_{k}$ is the probability for a helium droplet to pick up
$k$ SiO molecules, $P_\mathrm{I}$ is the probability to ionize a
droplet, and $P_\mathrm{CT}$ represents the probability for the
charge transfer from the droplet to its dopant. When the partial
evaporation of the droplet, that occurs at each pick up event, can
be neglected, the probability $P_{k}$ can be estimated by a Poisson
distribution \citep{Lewerenz95,Lewerenz97}. Since we want to
determine the smallest droplet size for which the assembly of a
given product is possible, the evaporation of the droplet is a
critical parameter and the Poisson distribution is no longer valid.

We regard the probability to pick up $k$ SiO molecules as a
product of the probabilities of single collision events multiplied
by the probability of not picking up additional molecules. This is
given by
\begin{equation}\label{eqPk}
P_k=\mathrm{exp}\left(-n_\mathrm{SiO}\sigma_{k}l_{k}\right)\prod_{i=0}^{k-1}\left[1-\mathrm{exp}\left(-n_\mathrm{SiO}\sigma_{i}l_{i}\right)\right] ,
\end{equation}
where $n_\mathrm{SiO}$ is the number density of SiO
molecules in the pick-up region, $\sigma_i$ the pick-up
(geometrical) cross section of the droplet after picking up $i$ SiO
molecules, and $l_i$ the rest length of the pick-up region. This
approach will be discussed in more detail in a forthcoming paper.

Similarly, the probability for a helium droplet to collide with an
electron in the ionizer can be expressed by
\begin{equation}\label{eqPion}
P_\mathrm{I}=1-\mathrm{exp}\left(-n_\mathrm{el}\sigma_{k}l_\mathrm{ion}\right) ,
\end{equation}
where $n_\mathrm{el}$ is equal to the number density of electrons
in the ionization region of the length $l_\mathrm{ion}$. The parameters $n_\mathrm{SiO}$
and $n_\mathrm{el}$ are determined by the conditions of the experiment such
as the temperature of the oven used to produce SiO molecules and the
current of the ionizer.

Finally, using the experimental data from \citet{Ruchti98}, we
conclude that the probability for charge transfer can be described
by a decreasing linear function on the number of He atoms in the
droplet. Hence,
\begin{equation}\label{eqCT}
P_\mathrm{CT}=1-cN_k .
\end{equation}
Initially, the parameter $c$ was set to achieve a probability of
50{\%} for the charge transfer in the largest helium droplets used
in our experiment. Moreover, we neglected the change of the helium
droplet size upon pick-up of SiO molecules. After a first evaluation
of the reaction energies (see below), we used the $\sigma_i$ values
determined by the experiment in Equation \ref{eqPk} and repeated the
analysis. This procedure was repeated until the derived binding
energies did not change anymore. It was also found that the
variation of the parameter $c$, which results in a change of the
charge transfer probability in the largest helium droplets from
10{\%} to 100{\%}, does not affect the values of the reaction
energies.

In this analysis, several approximations were used. We did not
consider the scattering of the helium droplet beam and the
non-symmetrical size distribution of the helium droplets. Moreover,
the velocity distributions of the SiO molecules and electrons was
not taken into account. All this is expected to cause
some inaccuracy of the calculated ion signal $P$. The effect
of such simplifications was already analyzed in a former paper and
the error was amounted to be up to 30\% for the evaluation of the
ion signals \citep{Vongehr10}. However, when the helium droplet is
below its minimum size required for the assembly of the desired
cluster, the difference between the measured and calculated ion
signal should be extremely large. Therefore, a simple
modeling of the ion signal dependence on the He droplet size allows
us to accurately identify the point where the experimental values
start to deviate from the calculated behavior. As a result, we
expect that such a procedure can provide an accuracy much better
than 30\% for the evaluation of the reaction energies. To
compare the theoretical and experimental values, we have scaled the
experimentally measured ion intensities by dividing by the
probability $P$ (see Equation~\ref{eqP}).

The experimental determination of reaction energies relies
on studies of He droplets, which were the subject of numerous
publications. Temperatures, sizes, and size distributions of He
droplets are well known \citep{Toennies04}. Excited molecules inside
He droplets relax to the ground state in a sub-nanosecond
time scale transferring their energy to the liquid helium, which is
the basis of our experimental analysis. The only other way to
dissipate the energy is a non-thermal ejection of reaction products
or photon emission. Non-thermally ejected reaction products cannot
be detected by mass spectrometry. The reason is the much smaller
ionization cross section compared with the ionization cross section of
He droplets. Our system does not detect these products and
therefore, they do not contribute to the determined reaction energy.
Photon emission can dissipate a considerable amount of energy
resulting in a strong discrepancy between experimentally and
computationally derived values of reaction enthalpy, which is not
the case in our analysis.

The scaled intensities [see Figure~\ref{fig6}(b)] show regimes with
different slopes. In the range of small helium droplets, the
capacity of the helium droplets to dissipate the heat of
condensation is not sufficient to allow the assembling of the
corresponding (SiO)$_k$ cluster. Non-scaled ion signals demonstrate
a very slow increase in this size regime [Figure~\ref{fig6}(a)],
which is mainly due to the size distribution of the helium droplets
and the presence of large helium droplets in the beam. However, this
increase is very small and, after scaling, we obtain a strong
decrease of the ion intensities as a function of the average droplet
size [Figure~\ref{fig6}(b)]. When the helium droplets reach the
minimum size, which is sufficient to assemble a (SiO)$_k$ cluster,
we observe a sudden change of the slope of the scaled signals. After
this point, the non-scaled ion intensities increase as predicted for
a short range and then start to rise. The rise is due to the fact
that (SiO)$_k$ ions are produced by fragmentation of larger clusters
during the ionization process. This gives an additional rise to the
ion signals on the (SiO)$_k$ mass.

To compute the condensation or reaction energies, we assume that the
evaporation of a single helium atom removes 5 cm$^{-1}$ equivalent
energy from the droplet \citep{Chin95}. The sizes of helium droplets
found from Figure~\ref{fig6} were used to calculate the energy of
the (SiO)$_k$ + SiO reactions. We calculate the difference between
the minimum numbers of helium atoms per droplet required to assemble
(SiO)$_k$ and (SiO)$_{k+1}$ clusters. An additional energy equivalent
to the evaporation of 1100 helium atoms is subtracted, due to the
pick-up of the last SiO molecule, with which the (SiO)$_{k+1}$
cluster was formed.

\end{document}